\documentclass{iau}
\usepackage{graphicx}
\pdfoutput=1

\title[Dziembowski scientific biography] 
{Wojtek Dziembowski: selected photos and \\ events of his scientific biography}

\author[Alexey A. Pamyatnykh]   
{Alexey A. Pamyatnykh}

\affiliation{Nicolaus Copernicus Astronomical Center,\\
Bartycka 18, 00-716, Warsaw, Poland \\ email: {\tt alosza@camk.edu.pl}}

\pubyear{2014}
\volume{301}  
\pagerange{1--8}
\jname{Precision Asteroseismology}
\editors{J.A. Guzik, W.J. Chaplin, G. Handler \& A. Pigulski, eds.}
\begin{document}

\maketitle

\begin{abstract}
Selected events of the scientific biography of Wojtek Dziembowski are briefly described,
and several related photos are presented. The full version of the presentation is available at
the IAU Symposium 301 webpage.
\keywords{biographies, Sun: helioseismology, stars: oscillations}
\end{abstract}


Wojtek Dziembowski was born in Warsaw in 1940 into the family of a
lawyer. In 1943 the family moved to the village Zawoja in Southern
Poland, and later, after the end of World War II, to the small town
\.Znin about 250 km to the north-west from Warsaw. Already in
primary school, being motivated by popular books of Sir James H.
Jeans, Wojtek decided to be an astronomer. In 1962 he obtained an
MSc degree in astronomy from the Jagellonian University in Krak\'ow.
On the basis of his MSc Thesis, Wojtek published  (just 50 years
ago!) his first paper, ``On the equations of internal constitution
of components in close binaries'' (\cite[Dziembowski
1963]{Dziem63}).


\begin{figure}[h]
\begin{center}
 \includegraphics[width=3.5in]{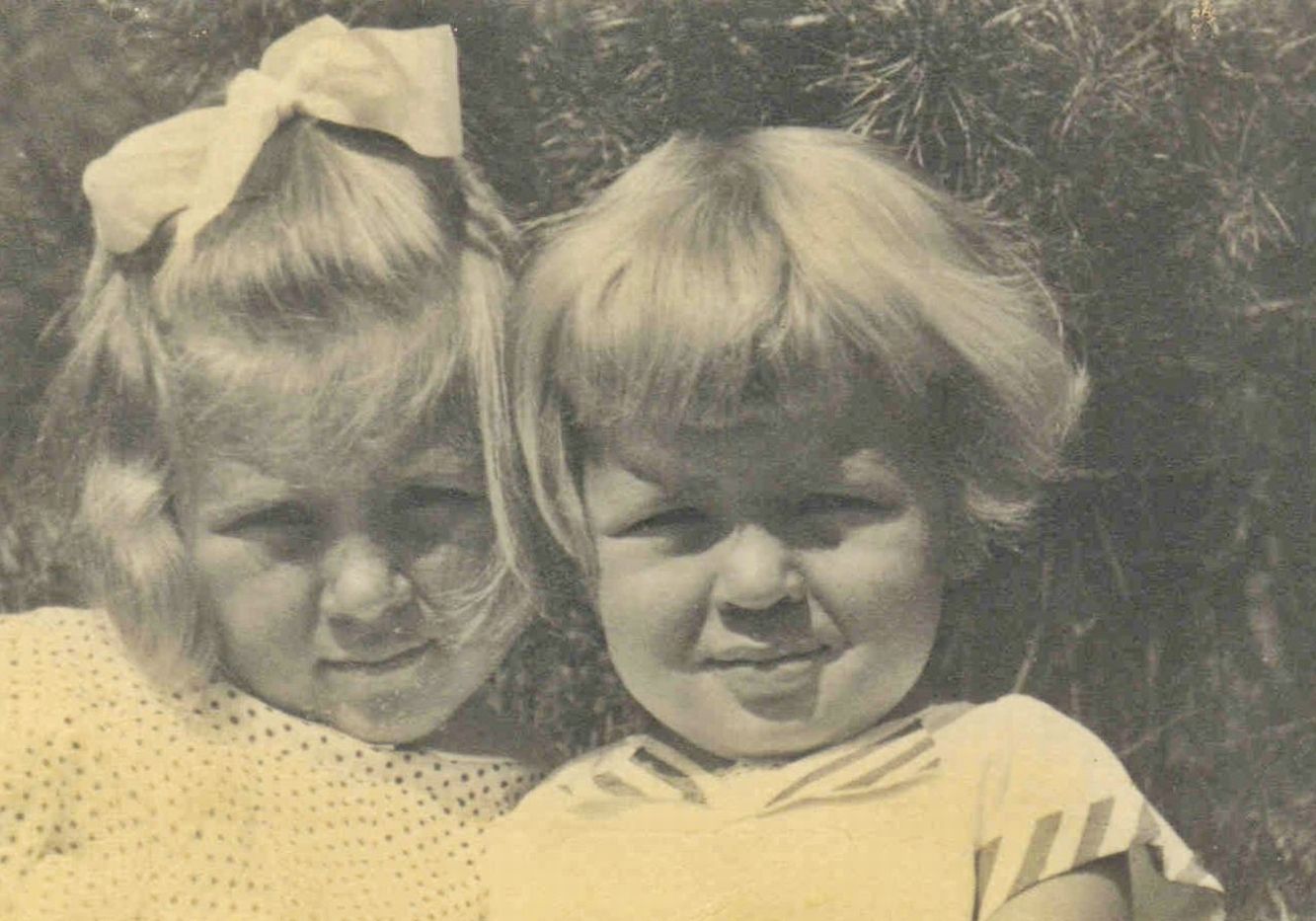}
 \caption{Wojtek and his older sister Teresa (Wojtek has three sisters) in
Zawoja, a village in Southern Poland, 1943. Photo from Archive of WAD.}
  \label{fig01}
\end{center}
\end{figure}


The next part in Wojtek{'}s career was played by Prof. Paul Ledoux (!) who visited Poland in 1963
and whom Wojtek told about his scientific work. Paul Ledoux wrote about Wojtek to Prof. Stefan
Piotrowski who invited Wojtek for graduate studies at Warsaw University.
In 1967 Wojtek defended his PhD thesis on tides in binary systems and, a few months later,
thanks to the recommendation of Prof. J\'ozef Smak, he obtained a postdoctoral fellowship at Columbia University,
New York, where he worked until 1969 with Prof. Norman Baker.
The main goal of their studies was an attempt to find an excitation mechanism for the $\beta$ Cephei
pulsations. Wojtek was the first to develop a computer code for linear nonadiabatic nonradial
oscillations. He obtained instability for realistic models of some classical variables
like $\delta$ Scuti stars but not for the $\beta$ Cephei-type variables. The results were never published,
and much later, at IAU Symposium 162 in France in 1993, Wojtek wrote in retrospect:
{\it ``For many years, explaining
the cause of $\beta$ Cephei variability has been a major challenge to stellar pulsation theory.
The problem is now solved, but we owe the solution
to progress in opacity calculation and not to new astrophysical ideas.  In fact, if
the OPAL opacities were available, the problem would have been solved many years ago
(Baker \& Dziembowski 1969, unpublished)''} (\cite[Dziembowski 1994]{Dziem94}).
It is absolutely fair that just Wojtek was among the first who really solved the puzzle of
the $\beta$ Cephei pulsations (\cite[Moskalik \& Dziembowski 1992]{MosDziem92})!


\begin{figure}[h]
\begin{center}
 \includegraphics[width=3.0in]{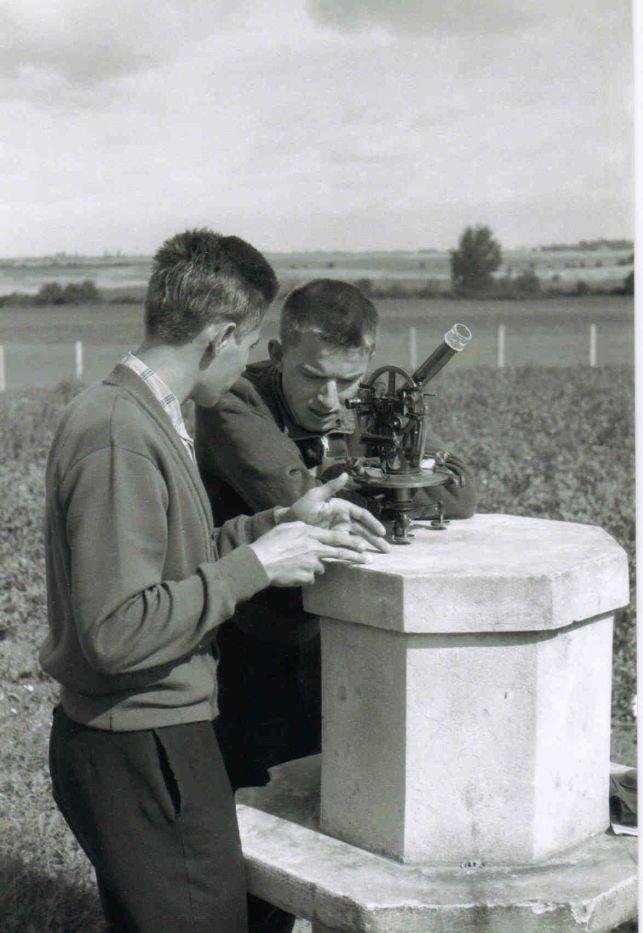}
 \caption{Wojtek and Bohdan Paczy\'nski at the summer astronomical practice,
Observatory Piwnice near Toru\'n, July 1960. Photo: M. G\'orski.}
   \label{fig02}
\end{center}
\end{figure}


From 1969 to the present, Wojtek has been working at Nicolaus Copernicus Astronomical Center (CAMK) in Warsaw,
and, since 1997, also at the Astronomical Observatory of the Warsaw University. In 1977 he defended
his Habilitation thesis on theoretical aspects of nonradial stellar oscillations,
and in $1978-1979$ he was working as Visiting Professor at the University of Arizona. In 1988 Wojtek obtained
the title of Professor, since 1989 he has been a Corresponding Member, and since 2007 a Member of the
Polish Academy of Sciences. Since 1997 he is also the Corresponding member of the Polish Academy of Learning, an elite scientific corporation of the Polish scientists. In $1987-1992$ Wojtek was the Director
of the CAMK, and in $2003-2006$ he was the President of IAU Commission 35 ``Stellar Constitution''.


\begin{figure}[h]
\begin{center}
 \includegraphics[width=4.5in]{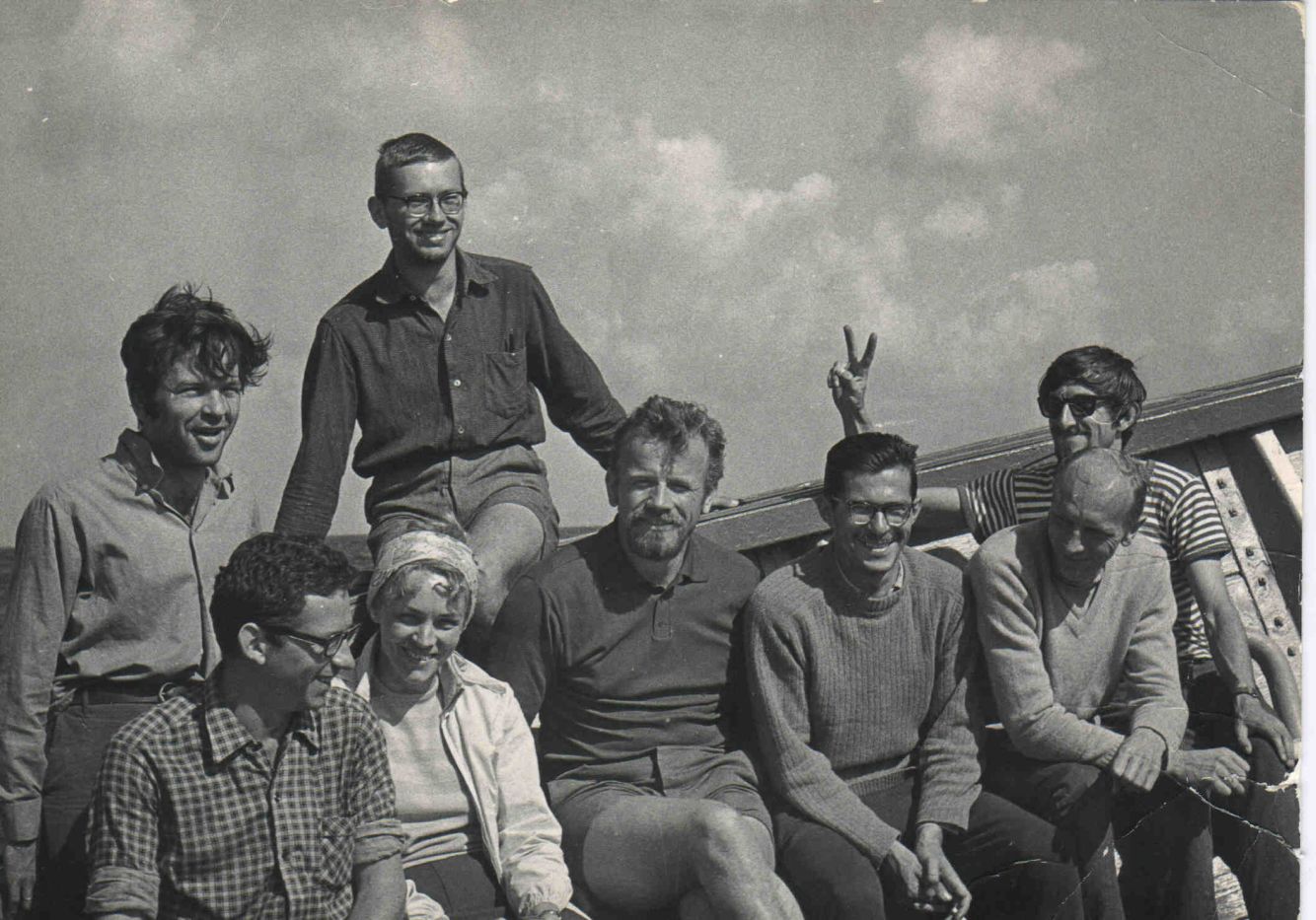}
 \caption{Onboard the motor yacht ``Podhalanin'' during a vojage of Polish astronomers
to the 14th IAU General Assembly in Brighton, August 1970. From the
left to the right: Wojtek Dziembowski, Janusz Zieli\'nski, Barbara
Ko{\l}aczek, Janusz Zi\'o{\l}kowski, 2nd officer Konstanty Pelak,
Mike Jerzykiewicz, Jan Bieniewski, and Adam Spodenkiewicz. Photo
from Archive of WAD.}
   \label{fig03}
\end{center}
\end{figure}


According to the SAO/NASA Astrophysics Data System (ADS, as of August 2013), Wojtek has 244 scientific
publications with total number of citations equal to about 6800.
Sixteen publications have each more than 100 citations, and Wojtek{'}s Hirsch index is equal to 44.
He is sole author of 52 publications; most of them are invited talks at many astrophysical meetings -
due to obvious restrictions for conference proceedings, they are relatively short
but written extremely carefully, clearly and precisely. As an example, I would like to refer again
to \cite[Dziembowski (1994)]{Dziem94}, where a very short and clear description of the $\kappa$-mechanism
is given with special attention to B-star pulsations. At that time Wojtek and collaborators
published results of detailed studies of $\beta$~Cephei and SPB pulsations
(\cite[Dziembowski \& Pamyatnykh 1993]{DziemPam93};
\cite[Dziembowski et al.~1993]{DMP93}).

Note that two of Wojtek{'}s single-author papers have been published only recently -- on dipolar modes in red giants
(\cite[Dziembowski 2012a]{Dziem12a}) and on puzzling frequencies in first-overtone Cepheids
(\cite[Dziembowski 2012b]{Dziem12b}).

A few of Wojtek{'}s papers have become classics, as they gave rise to the creation of new directions
for theoretical studies of stellar pulsations and for interpretations of observations of multiperiodic
variables. Wojtek was one of the first who developed detailed algorithms and codes for computation of
linear nonadiababic nonradial stellar oscillations (\cite[Dziembowski 1971, 1977a]{Dziem71}),
a method of nonradial mode identification from observations of light and radial velocity variations
(\cite[Dziembowski 1977b]{Dziem77b}), the theory of nonlinear mode coupling in oscillating
stars (\cite[Dziembowski 1982]{Dziem82}), a method of taking into account effects of differential rotation
on stellar oscillations (\cite[Dziembowski \& Goode 1992]{DziemGoode92}), an algorithm for numerical
solution of the inverse problem in helioseismology
(\cite[Dziembowski et al.~1990]{DPS90};
see the contribution by Douglas Gough in this volume for Wojtek{'}s many other achievements in helioseismology).

A distinctive property of Wojtek{'}s main papers is that they became most popular and cited
more than 20 years after publication -- this means that these papers were ahead of their time
and they predicted or even defined the direction of stellar pulsation studies!
For example, Wojtek{'}s most cited paper on light and radial velocity variations
in a nonradially oscillating star (\cite[Dziembowski 1977b]{Dziem77b}) with a total number
of 221 citations has been cited 42 times in 1978--1987 and 82 times in 2003--2012.
Indeed, owing to recent cosmic missions and high-precision ground-based observations, a lot of
multiperiodic variables have been detected, and elaboration of the methods of nonradial mode identifications
is of highest importance for asteroseismology. Jagoda Daszy\'nska-Daszkiewicz and Wojtek,
in collaboration with other colleagues, continue to refine his method,
taking into account effects of rotation (see \cite[Daszy\'nska-Daszkiewicz et al.~2002]{DDPG02}
and later papers).


\begin{figure}[h]
\begin{center}
 \includegraphics[width=4.0in]{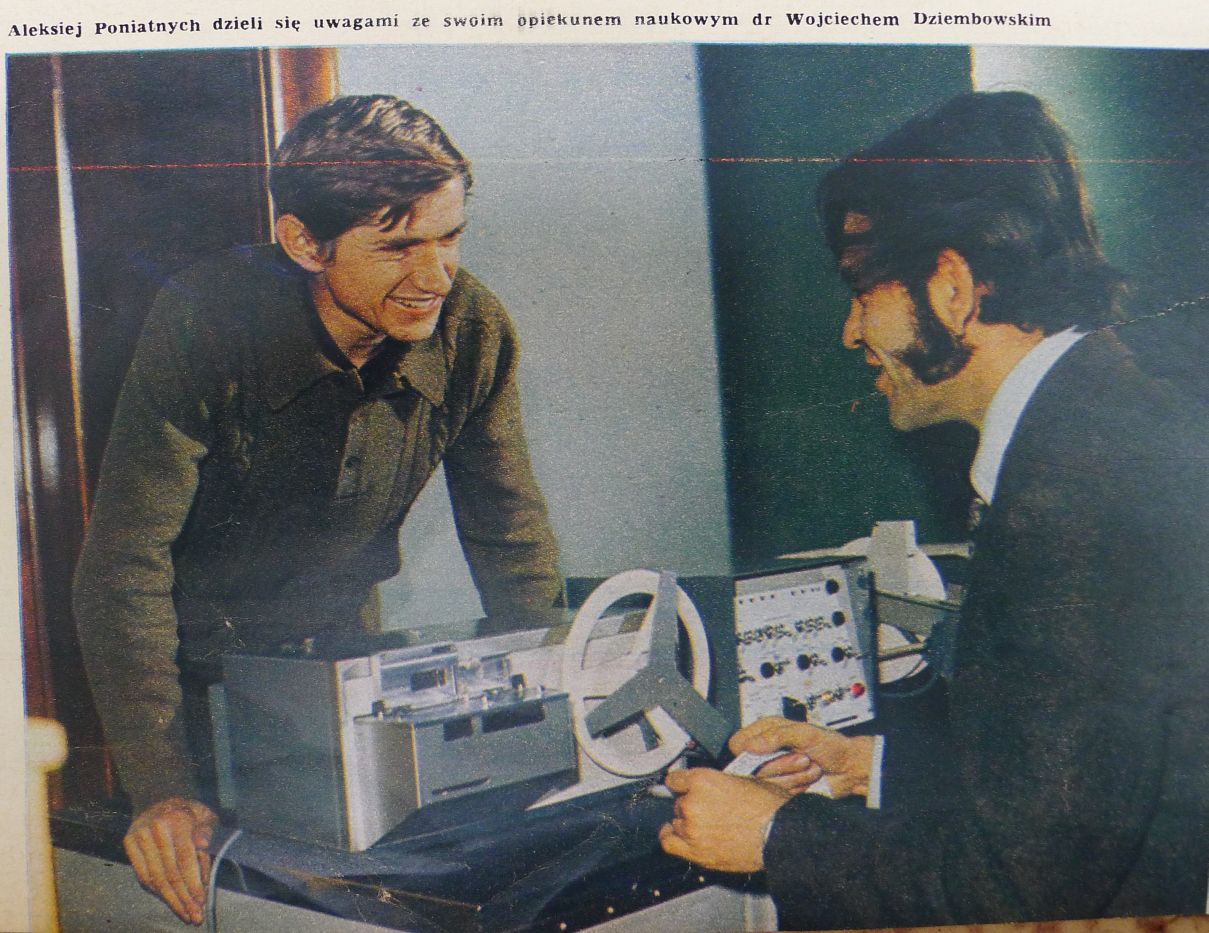}
 \caption{``Alexey Pamyatnykh shares his considerations with his scientific adviser
Dr Wojciech Dziembowski'', Warsaw,  Autumn 1972.
(Computer ODRA 1204, read-punch unit for paper tape with an Algol program.)
Photo and caption from the magazine ``Przyja\'z\'n'' (``Friendship''), January 1973.
}
   \label{fig04}
\end{center}
\end{figure}


Many of Wojtek{'}s investigations have been performed in collaboration with colleagues both from Poland and from other countries. In my opinion, all of Wojtek{'}s collaborators are happy
to work with him; they  get creative scientific impulses from his comprehensive knowledge
and from his fantastic intuition.

Wojtek was the scientific adviser of 7 students who successfully defended their PhD theses
(Ryszard Sienkiewicz, Zbigniew Loska, Pawe{\l} Moskalik, Ma{\l}gorzata Kr\'olikowska,
Pawe{\l} Artymowicz, Jan Zalewski, Rafa{\l} Nowakowski).

When one considers 225 of Wojtek{'}s publications with less than 6 authors, his closest collaborators appear
to be Philip Goode (as coauthor of 48 publications), Alexey Pamyatnykh (40),
Jagoda Daszy\'nska-Daszkiewicz (16), Ryszard Sienkiewicz (15), Marie-Jo Goupil (11)
and Aleksander Kosovichev (11). It can be undoubtedly stated that he is the creator of the Polish school of
asteroseismology, and now his informal group includes researchers from Warsaw and Wroc{\l}aw,
and collaborates with many other scientists around the world (I would like to note that one of the most
active researchers of pulsating stars, Gerald Handler, is now working at the Copernicus Center in Warsaw.)


\begin{figure}[h]
\begin{center}
 \includegraphics[width=4.0in]{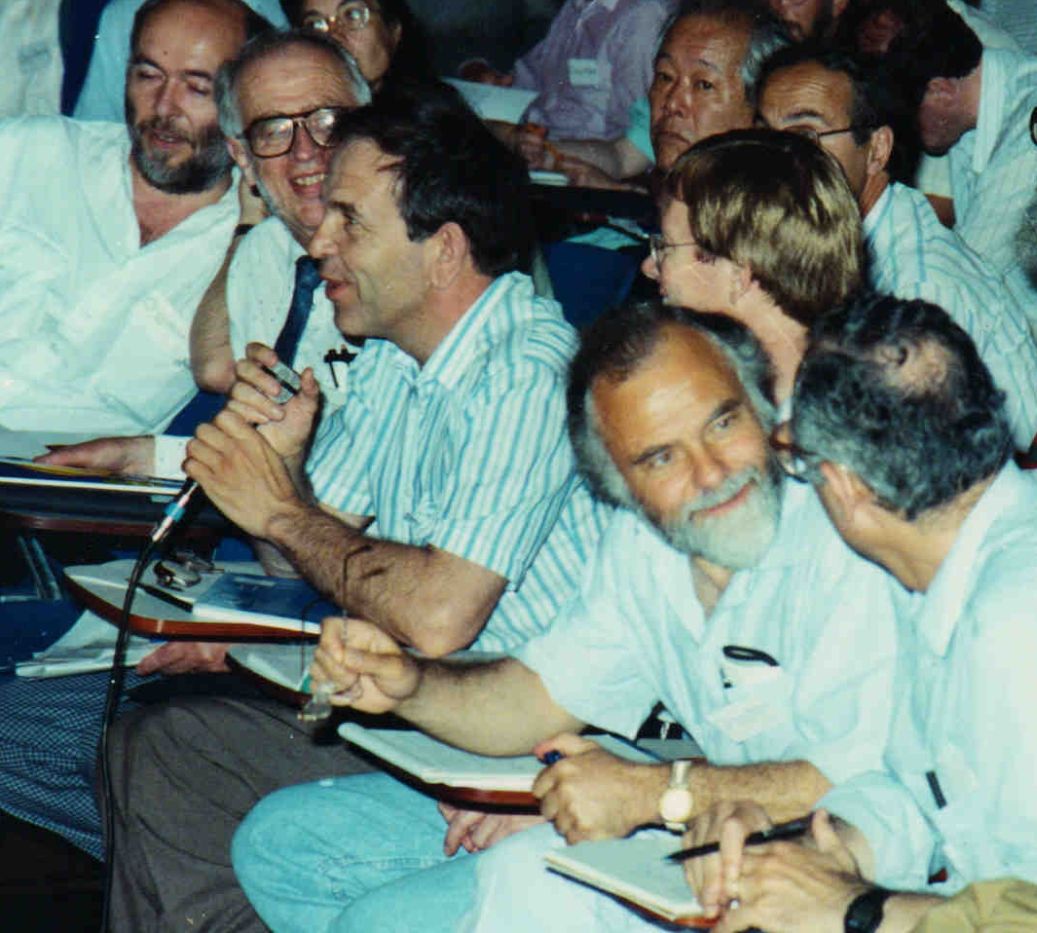}
 \caption{With Arthur Cox, Douglas Gough, Philip Goode and other colleagues,
IAU Coll.~121 ``Inside the Sun'', Versaille,  May 1989.
Wojtek and Philip Goode delivered a talk
``Magnetic field in the Sun{'}s interior from oscillation data''.
From Archive of WAD.}
   \label{fig05}
\end{center}
\end{figure}


\begin{figure}[h]
\begin{center}
 \includegraphics[width=4.0in]{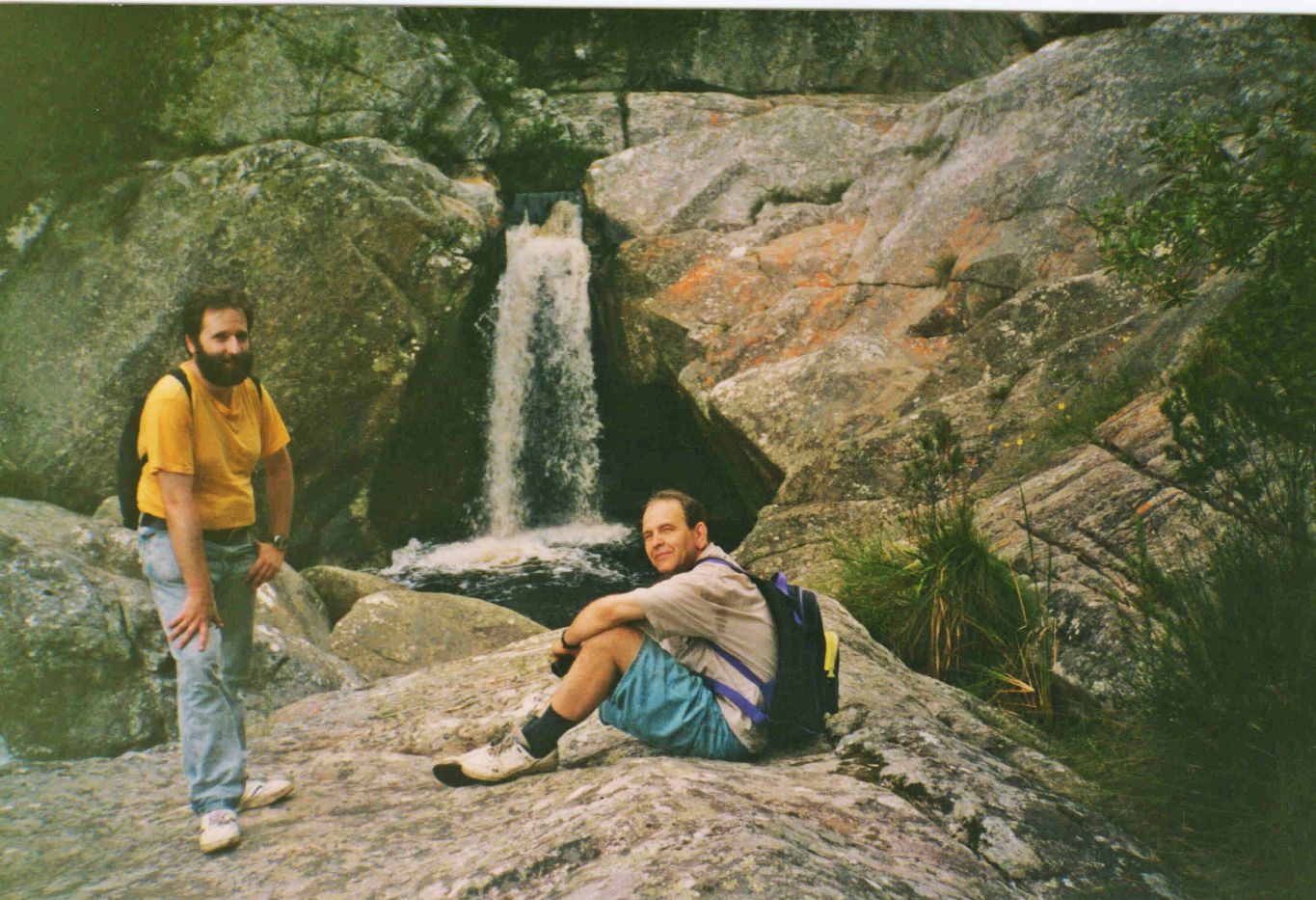}
 \caption{With Pawe{\l} Moskalik after the end of IAU Coll. 155,
``Astrophysical Applications of Stellar Pulsation'', February 1995, Cape Town.
At the conference, Wojtek with coauthors had four presentations, and Pawe{\l} presented
the up-to-date results of our group on pulsating OB-stars. From Archive of P.M.}
   \label{fig06}
\end{center}
\end{figure}


\begin{figure}[h]
\begin{center}
 \includegraphics[width=4.5in]{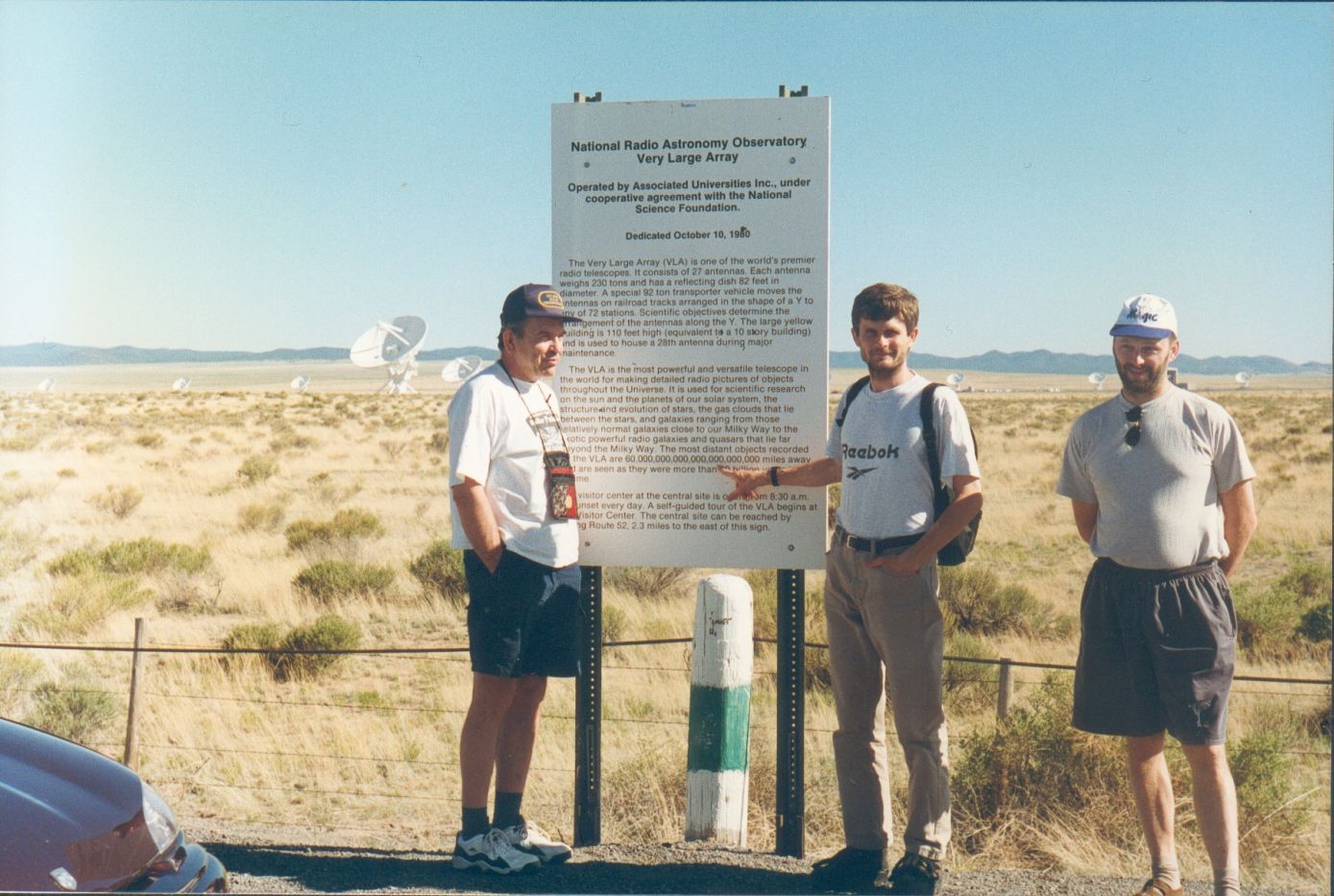}
 \caption{On the way to the Los Alamos Meeting ``A Half Century of Stellar Pulsation
Interpretations. A Tribute to Arthur N. Cox'',
Wojtek with Jurek Krzesi\'nski and Andrzej Pigulski.
New Mexico,  June 1997 (near VLA of NRAO).
At the conference, Wojtek gave an introductory talk on helio- and asteroseismology.
Photo: A. Pamyatnykh.}
   \label{fig07}
\end{center}
\end{figure}


\begin{figure}[h]
\begin{center}
 \includegraphics[width=4.2in]{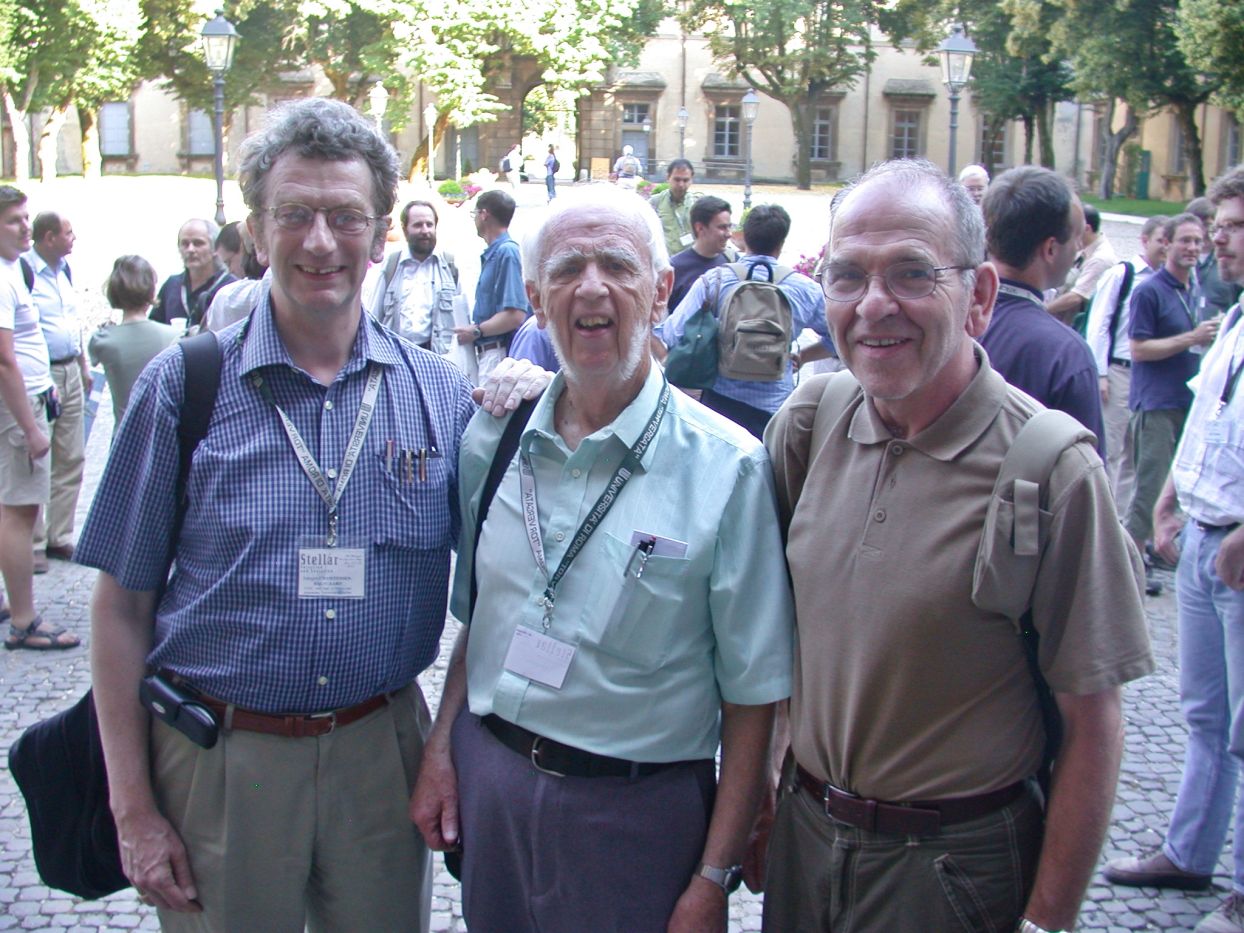}
 \caption{Conference ``Stellar Pulsation and Evolution'', Monte Porzio Catone, Italy, June 2005.
J{\o}rgen Christensen-Dalsgaard, Arthur Cox and Wojtek. In his introductory talk
on asteroseismology, Wojtek gave examples of interpretations of rich oscillation spectra of
different stars, including solar-like, sdB and $\beta$~Cep-type pulsators. Photo: J.A. Guzik}
   \label{fig08}
\end{center}
\end{figure}


\begin{figure}[h]
\begin{center}
 \includegraphics[width=4.0in]{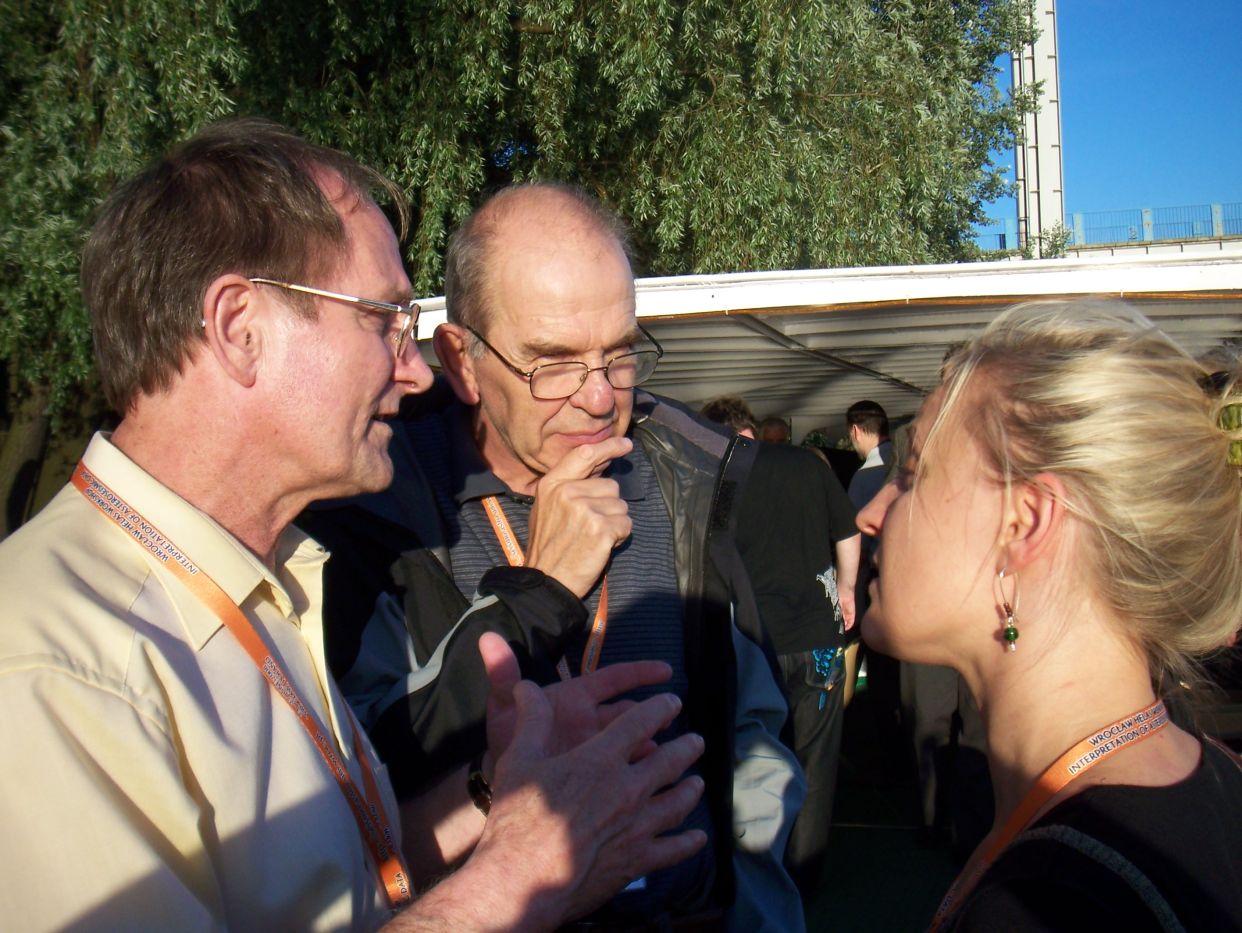}
 \caption{With Jagoda Daszy\'nska-Daszkiewicz and Mike Breger.
The HELAS Workshop ``Interpretation of Asteroseismic Data'',
Wroc{\l}aw, June 2008. Photo: A. Pamyatnykh.}
   \label{fig09}
\end{center}
\end{figure}


\begin{figure}[h]
\begin{center}
 \includegraphics[width=4.5in]{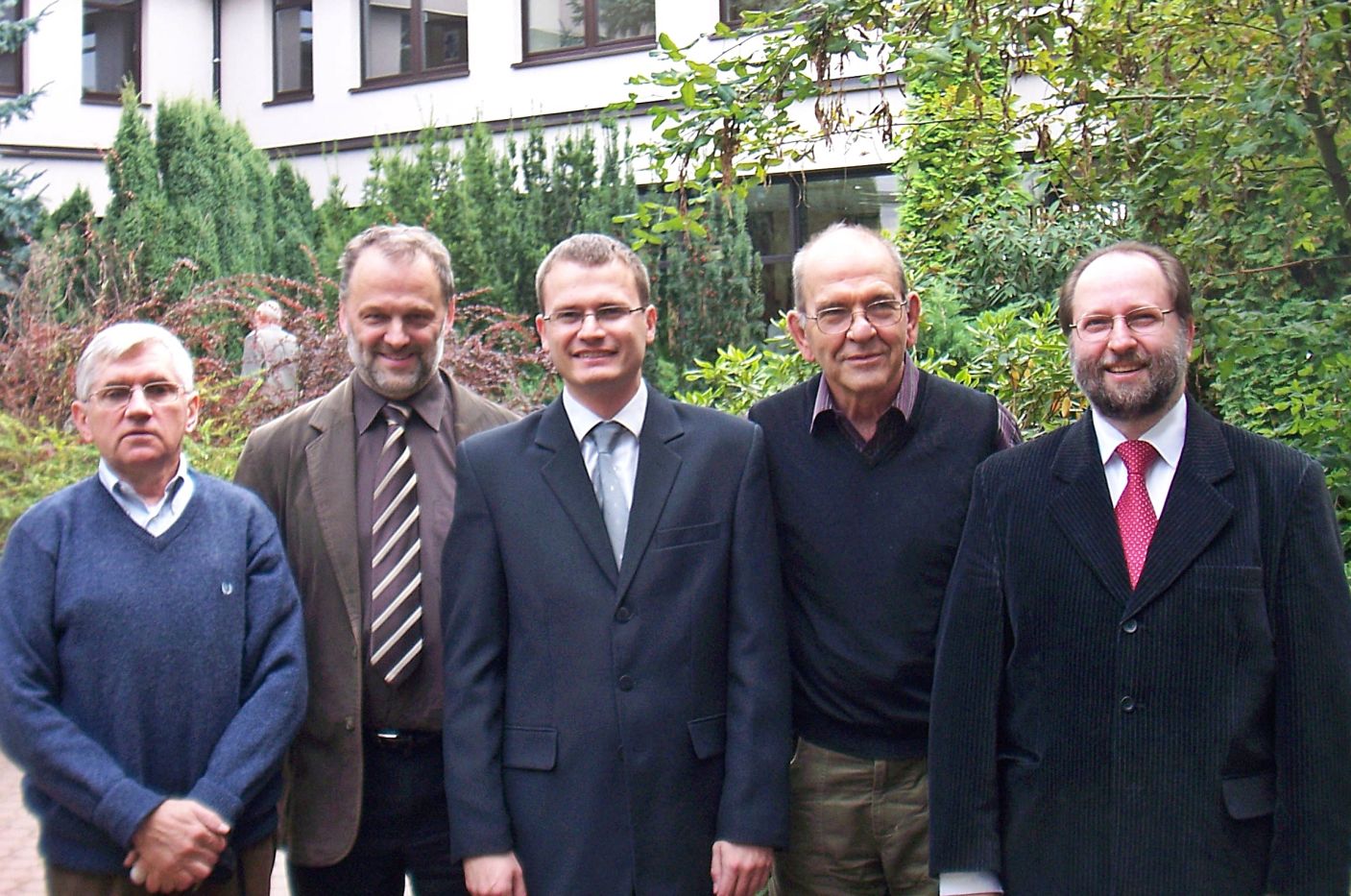}
 \caption{From right to left:
Pawe{\l} Moskalik,  Wojtek Dziembowski,  Radek Smolec,  Andrzej Pigulski,  Alexey Pamyatnykh.
CAMK, Warsaw, September 2009, Defense of PhD dissertation by Radek Smolec. Pawe{\l} was the
scientific adviser of Radek, and  Wojtek was many years ago the scientific adviser of Pawe{\l}.
This is a part of our group headed by Wojtek and now also by Gerald Handler.}
   \label{fig10}
\end{center}
\end{figure}


\begin{figure}[h]
\begin{center}
 \includegraphics[width=4.5in]{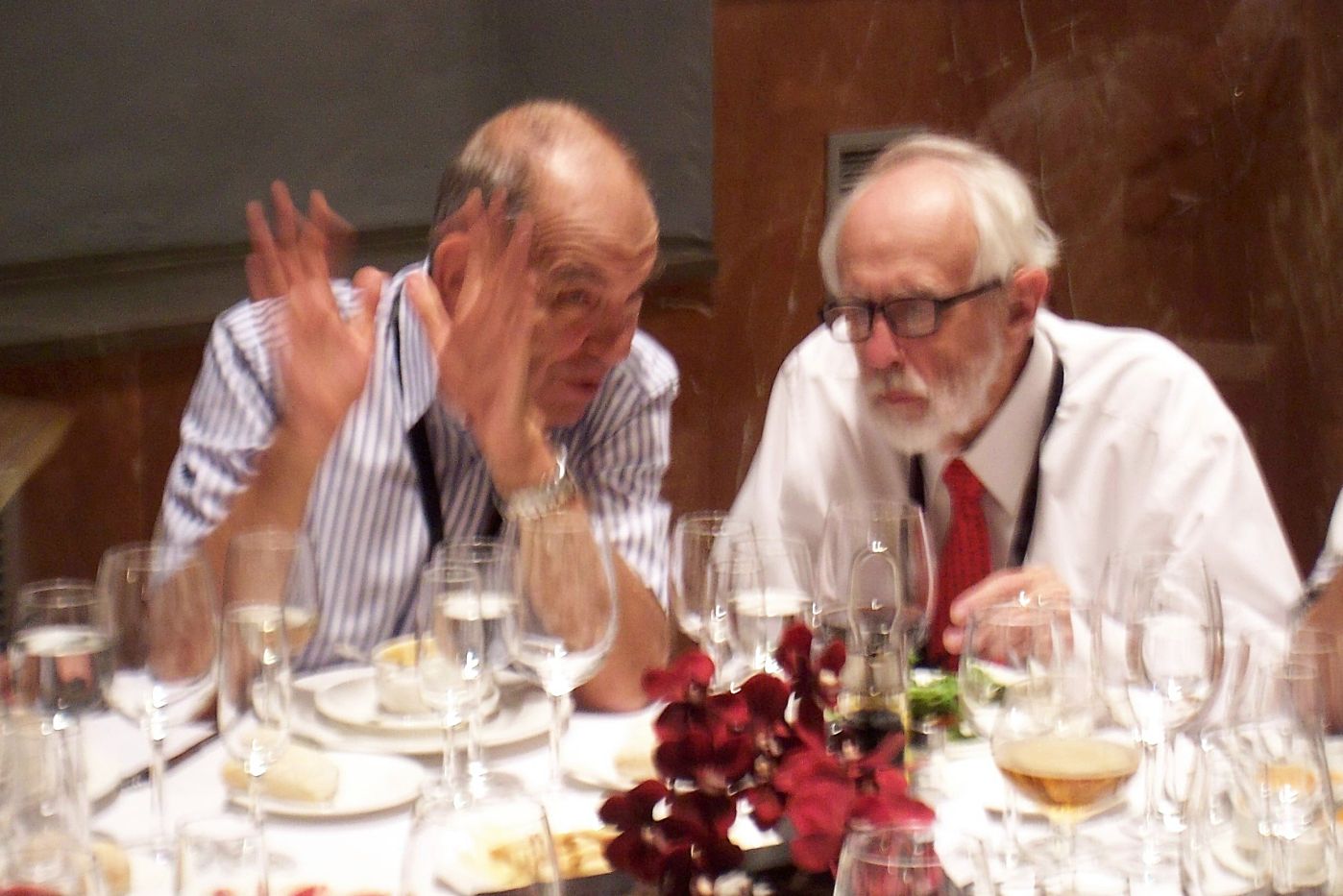}
 \caption{Conference ``Impact of new instrumentation and new insights in stellar pulsations'',
Granada, September 2011. With Arthur Cox at the conference dinner. Photo: A. Pamyatnykh.}
   \label{fig11}
\end{center}
\end{figure}


\acknowledgements I acknowledge partial financial support from the Scientific Organizing Committee
and from the Polish NCN grant 2011/01/B/ST9/05448. I am grateful to Wojtek Dziembowski,
Joyce Guzik, Mike Jerzykiewicz and Pawe{\l} Moskalik for the photos which have been used
in this contribution and presented at the Symposium.


The extended version of the presentation at the Symposium is available at\\
http://www.astro.uni.wroc.pl/IAUS301\_Talks/Day1-0950-Pamyatnykh.pdf.


\begin{thebibliography}{}

\bibitem[Daszy\'nska-Daszkiewicz et al.~(2002)]{DDPG02}
{Daszy\'nska-Daszkiewicz, J., Dziembowski, W.A., Pamyatnykh, A.A., \& Goupil, M.-J.} 2002,
\textit{A\&A}, 392, 151

\bibitem[Dziembowski (1963)]{Dziem63}
{Dziembowski, W.A.} 1963,
\textit{AcA}, 13, 157

\bibitem[Dziembowski (1971)]{Dziem71}
{Dziembowski, W.A.} 1971,
\textit{AcA}, 21, 289

\bibitem[Dziembowski (1977a)]{Dziem77a}
{Dziembowski, W.A.} 1977a,
\textit{AcA}, 27, 95

\bibitem[Dziembowski (1977b)]{Dziem77b}
{Dziembowski, W.A.} 1977b,
\textit{AcA}, 27, 203

\bibitem[Dziembowski (1982)]{Dziem82}
{Dziembowski, W.A.} 1982,
\textit{AcA}, 32, 147

\bibitem[Dziembowski (1994)]{Dziem94}
{Dziembowski, W.A.} 1994, in: L.A. Balona, H.F. Henrichs, \& J.M. Le Contel (eds.),
\textit{Pulsation, rotation and mass loss in early-type stars}, Proc.~IAU Symp.~162, 55

\bibitem[Dziembowski (2012a)]{Dziem12a}
{Dziembowski, W.A.} 2012a,
\textit{A\&A}, 539, A83

\bibitem[Dziembowski (2012b)]{Dziem12b}
{Dziembowski, W.A.} 2012b,
\textit{AcA}, 62, 323

\bibitem[Dziembowski \& Goode (1992)]{DziemGoode92}
{Dziembowski, W.A., \& Goode, P.R.} 1992,
\textit{ApJ}, 394, 670

\bibitem[Dziembowski \& Pamyatnykh (1993)]{DziemPam93}
{Dziembowski, W.A., \& Pamyatnykh, A.A.} 1993,
\textit{MNRAS}, 262, 204

\bibitem[Dziembowski, Pamyatnykh \& Sienkiewicz (1990)]{DPS90}
{Dziembowski, W.A., Pamyatnykh, A.A., \& Sienkiewicz, R.} 1990,
\textit{MNRAS}, 244, 542

\bibitem[Dziembowski, Moskalik \& Pamyatnykh (1993)]{DMP93}
{Dziembowski, W.A., Moskalik, P., \& Pamyatnykh, A.A.} 1993,
\textit{MNRAS}, 265, 588

\bibitem[Moskalik \& Dziembowski (1992)]{MosDziem92}
{Moskalik, P., \& Dziembowski, W.A.} 1992,
\textit{A\&A}, 256, L5

\end{thebibliography}
\end{document}